\documentclass[prd,twocolumn,amsmath,amssymb,floatfix, superscriptaddress,nofootinbib]{revtex4}

\usepackage{bm}
\usepackage{amsmath}
\usepackage{epsf}
\usepackage{color}
\usepackage{natbib}
\usepackage{graphicx}
\usepackage{hyperref}
\usepackage{ifthen}
\usepackage{xstring}

\newcommand{\be}{\begin{equation}}
\newcommand{\ee}{\end{equation}}
\newcommand{\bea}{\begin{eqnarray}}
\newcommand{\eea}{\end{eqnarray}}
          
\newcommand{\refeq}[1]{Eq.~(\ref{eqn:#1})}          
          
\newcommand{\reffig}[1]{Fig.~\ref{fig:#1}}          
\newcommand{\reftab}[1]{Tab.~\ref{tab:#1}}

\newcommand{\refsec}[1]{Section~\ref{sec:#1}}

\newcommand{\Msunh}{\,M_{\odot}/h}

\newcommand{\Om}{\Omega_m}

\newcommand{\Mxeff}{M_{\rm eff}}
\newcommand{\seff}{\sigma_8^{\rm eff}}
\newcommand{\fRzero}{$|f_{R0}|=10^{-4}$ }
\newcommand{\LCDM}{{\Lambda{\rm CDM}}}

\newcommand\Mest{\ensuremath{M^{\text{est}}}}

\newcommand\Mmin{\ensuremath{M_{\text{min}}}}
\newcommand\Mmax{\ensuremath{M_{\text{max}}}}
\newcommand\zmin{\ensuremath{z_{\text{min}}}}
\newcommand\zmax{\ensuremath{z_{\text{max}}}}

\begin{document}
\title{Cluster Constraints on $\bm{f(R)}$ Gravity}

\author{Fabian Schmidt}
\affiliation{Kavli Institute for Cosmological Physics, Department of Astronomy \& Astrophysics, and Enrico Fermi Institute, University of Chicago, Chicago IL 60637, U.S.A.}
\author{Alexey Vikhlinin}
\affiliation{Harvard-Smithsonian Center for Astrophysics, 60 Garden
  St., Cambridge MA 02138, USA, and Space Research Institute, Moscow, Russia}
\author{Wayne Hu}
\affiliation{Kavli Institute for Cosmological Physics, Department of Astronomy \& Astrophysics, and Enrico Fermi Institute, University of Chicago, Chicago IL 60637, U.S.A.}

\date{\today}

\begin{abstract}
Modified gravitational forces in models that seek to explain cosmic acceleration without
dark energy typically predict deviations in the abundance of massive dark matter halos.
We conduct the first, simulation calibrated, cluster abundance 
constraints on a modified gravity model, specifically the
modified action $f(R)$ model.  The local cluster abundance, when combined with geometric and high redshift
data from the cosmic microwave background, supernovae, $H_0$ and baryon acoustic oscillations, improve previous constraints by nearly 4 orders of magnitude in the field amplitude.
These limits correspond to a 2 order of magnitude improvement in the bounds
on the range of the force modification from the several Gpc scale to the tens of Mpc scale.
 \end{abstract}

\maketitle

\section{Introduction} \label{sec:intro}

In $f(R)$ models, cosmic acceleration arises not from an exotic form of energy with negative pressure
but from a modification of gravity.   Here the Einstein-Hilbert action is 
augmented by a general function $f(R)$ of
the Ricci or curvature scalar $R$  \cite{Caretal03,NojOdi03,Capozziello:2003tk}.  
Such modifications not only can accelerate the background expansion but also
generically lead to enhancements in gravitational forces on small scales.

$f(R)$ gravity is equivalent to a scalar-tensor theory, where $f_R=df/dR$
is the additional scalar degree of freedom. This field has a mass and propagates
on scales smaller than the 
associated Compton wavelength. 
Well within the Compton wavelength, the scalar mediates a 4/3 enhancement of
 gravitational forces,
with corresponding strong effects on the growth of structure in the Universe.
These enhancements are quantified by the mass of the field or
equivalently by the value of the field  in the
background, $f_{R0}$.  
In order to pass stringent Solar System constraints, viable $f(R)$ models
employ the chameleon effect which allows the field to become very massive in 
high-density environments \cite{khoury04a}. However, in order for the chameleon
effect to become active, the background field should be smaller than
the typical depth of cosmological potential wells, $|f_{R0}| < |\Psi| \sim 10^{-6}-10^{-5}$
\cite{HuSaw07a}.

Independently of this Solar System constraint, however,
it is worth studying the constraints that can be placed  on $f(R)$ gravity
directly from cosmological observations. In the linear regime,
these provide only weak 
constraints.  Changes in the low multipole anisotropy of the
cosmic microwave background (CMB) only place order unity bounds on the value
of $f_{R0}$, while
changes to the shape of the matter power spectrum, though larger,
can be mimicked by galaxy bias \cite{SonPeiHu07}.   

The effect of enhanced forces can be substantially more prominent in the non-linear regime.  
Cosmological simulations have shown that for field values larger than $|f_{R0}| \sim 10^{-5}$
the abundance of rare massive halos are enhanced substantially
\cite{halopaper}.  Counts of galaxy clusters therefore provide the opportunity to improve
cosmological constraints on $f(R)$ models ultimately by 4-5 orders of magnitude. 

In this Paper, we quantify current cluster abundance constraints on $f(R)$ 
models from a combined sample of low-redshift X-ray clusters.  
We begin in \S \ref{sec:model} with a description of the model and its impact on cluster
predictions.   In \S \ref{sec:clusters} we describe the likelihood analysis of the 
local cluster abundance data.  We combine these constraints with geometric and
high redshift constraints in \S \ref{sec:combined} to obtain upper limits on the modification
to gravity. We discuss these results in \S \ref{sec:discussion}.

\section{$f(R)$  Cluster Abundance}

In this section, we describe the enhancement that $f(R)$ models make to the
cluster abundance.   We begin in \S \ref{sec:model} with a brief review of the $f(R)$
model itself.  We describe cosmological simulations with representative parameter
choices in \S \ref{sec:sim} which are used to calibrate mass function enhancements
across a wider range of parameters
in \S \ref{sec:dndm}.

\subsection{$f(R)$ Model}
\label{sec:model}

In the  $f(R)$ model, the Einstein-Hilbert action is augmented with a general function of the scalar curvature $R$,
 \begin{eqnarray}
S_{G}  =  \int{d^4 x \sqrt{-g} \left[ \frac{R+f(R)}{16\pi G}\right]}\,. 
\label{eqn:action}
\end{eqnarray}
Here and throughout $c=\hbar=1$.    
Gravitational force enhancements are associated
with an additional scalar degree of freedom $f_{R}\equiv df/dR$ and have a range
given by the Compton wavelength 
$\lambda_C= (3 d f_R/dR)^{1/2}$. This additional attractive force
leads to the enhancement in the abundance of rare massive dark matter halos described below.

For definiteness, we choose the functional form for $f(R)$ given in \cite{HuSaw07a} (with $n=1$):
\begin{equation}
f(R) = - 2\Lambda \frac{R}{R+\mu^2},
\end{equation}
with two free parameters, $\Lambda$, $\mu^2$. Note that as $R\rightarrow 0$,
$f(R)\rightarrow 0$, and hence the model does not contain a cosmological
constant. For $R \gg \mu^2$,  the function $f(R)$ can be approximated as
\begin{eqnarray}
f(R) = -2 \Lambda - f_{R0} \frac{\bar R_0^2}{ R} \,,
\label{eqn:fRapprox}
\end{eqnarray}
with $f_{R0}= -2\Lambda \mu^2/\bar R_0^2$ replacing $\mu$ as the second parameter
of the model.   
  Here we define $\bar R_{0}=\bar R(z=0)$, 
so that $f_{R0}= f_{R}(\bar R_{0})$, where overbars denote the quantities of the
background spacetime.  Note that if $|f_{R0}| \ll 1$ the curvature scales set by $\Lambda
={\cal O}(R_0)$
and $\mu^2$ differ widely and hence the $R \gg \mu^2$ approximation is valid today and for
all times in the past.

The background expansion history mimics $\Lambda$CDM with $\Lambda$ as
a true cosmological constant
to order $f_{R0}$.   Therefore in the limit 
$|f_{R0}| \ll 10^{-2}$, the $f(R)$ model and $\Lambda$CDM are essentially indistinguishable with geometric tests.    Nonetheless geometric tests do constrain one of the
two parameters ($\Lambda$) leaving the cluster abundance to constrain the
other parameter ($f_{R0}$) which controls the strength and range of the force modification. 
With the functional form of Eq.~(\ref{eqn:fRapprox}), the comoving Compton wavelength becomes
\begin{equation}
{\lambda_C \over 1+z}= \sqrt{ 6 | f_{R0}| {R_0^2 \over R^3}} \,,
\end{equation}
with a value today of 
\begin{equation}
\lambda_{C0} \approx 32 \sqrt{ | f_{R0}| \over 10^{-4}} {\rm Mpc} \,.
\end{equation}

The behavior of $f(R)$ gravity is described by the modified Einstein 
equations. Specifically, the trace of the linearized Einstein equations yields
the $f_R$ field equation
\begin{eqnarray}
\nabla^2 \delta  f_{R} = \frac{a^{2}}{3}\left[\delta R(f_{R}) - 8 \pi G \delta \rho_{\rm m}\right] \,,\label{eqn:frorig}
\end{eqnarray}
where time derivatives have been neglected 
compared with spatial derivatives, coordinates are comoving, $\delta f_R = f_R(R)-f_R(\bar R)$,
$\delta R = R -\bar R$, $\delta \rho_{\rm m} = \rho_{\rm m} - \bar \rho_{\rm m}$.
Note that the local curvature $R$ is given as a function of the local field
value $f_R$.
The time-time component  returns the modified Poisson 
equation
\begin{eqnarray}
\nabla^2 \Psi = {4\pi G}a^{2} \delta \rho_{\rm m} - \frac{1}{2} \nabla^2 \delta  f_{R} \,.\label{eqn:potorig}
\end{eqnarray}
Here $\Psi$ is the Newtonian potential or time-time metric perturbation $2\Psi = \delta g_{00}/g_{00}$ in the longitudinal gauge.    These two equations define a closed system for
the Newtonian potential given the density field.    The matter falls in the Newtonian potential
as usual and so the modifications to gravity are completely contained in the equation for $\Psi$.

Due to the conformal equivalence of $f(R)$ and scalar-tensor theories and the conformal
invariance of electromagnetism, the geodesics of photons are unmodified by the presence of the
scalar $f_R$ field save for conformal rescaling factors of $1+f_R$ \cite{BekSan94}.  This means that given a fixed density field, e.g. a halo
of mass $M$, the {\it lensing} potential will be identical to the one in GR in the $|f_R|\ll 1$ limit 
that we work in. In other
words, we will measure the ``true'' mass $M$ through lensing. 

On the other
hand, the mass inferred from dynamical measures, $M_{\rm dyn}$, such as velocities
and virial temperatures is related to the {\it dynamical} potential which
will be different in the presence of the $f_R$ field.
In the low curvature limit where $\delta R \ll 8\pi G \delta \rho_{\rm m}$, \refeq{potorig} shows
that the dynamical potential will be enhanced by $4/3$. Hence the mismatch
between $M_{\rm dyn}$ and $M$ could be as large as 33\%.
Conversely, field fluctuations can saturate in deep gravitational potentials
as  $f_R \rightarrow 0$.  Here $\delta R \approx 8\pi G \delta \rho_{\rm m}$ and force
modifications disappear via the chameleon mechanism. Then, if the whole mass
is contained in the saturated region, $M_{\rm dyn}=M$.   We discuss
the mass calibration of the cluster sample in \refsec{clusters}.

\subsection{Simulations}
\label{sec:sim}

We use a modified N-body simulation as described in \cite{Oyaizu:2008sr} and used 
in \cite{Oyaizu:2008tb,halopaper} to quantify the impact of the force modification on 
the cluster abundance.  
Specifically we employ the system of equations defined by the 
modified Poisson equation
(\ref{eqn:potorig}) and the $f_R$ field equation (\ref{eqn:frorig}) in the 
context of cosmological structure formation.
The field equation for $f_R$ is solved on a fixed cubic grid,
using a non-linear relaxation algorithm.
The potential $\Psi$ is computed from the density and $f_R$ fields
using the fast Fourier transform method.
The dark matter particles are then moved according to the gradient of the
computed potential, $-\nabla \Psi$.

The simulations were performed for a range of background field values 
$|f_{R0}|= 10^{-6}-10^{-4}$. We also simulated $|f_{R0}|=0$ which is
equivalent to $\Lambda$CDM, using the same initial conditions.  
Note that the background expansion history for all
runs is indistinguishable from $\Lambda$CDM to ${\cal O}(f_{R0})$.

Cosmological parameters were fixed in the simulations to a
flat  $\Omega_\Lambda=0.76$,
$\Omega_b=0.04181$, $H_0=73$ km/s/Mpc model and initial power in curvature 
fluctuations
$A_s=(4.82\times 10^{-5})^2$ at $k=0.02$Mpc$^{-1}$ with a tilt of $n_s=0.958$.
All simulations possess 512 grid cells in each direction and 
$N_p=256^3$ particles. 
Halos were identified in the simulations using a standard spherical overdensity
halo finder \cite{halopaper}.
In the next section, we describe our model
for the $f(R)$ effects on the halo mass function, which allows us
to extend predictions to a range of cosmological parameter values.

\subsection{Mass Function Enhancement}
\label{sec:dndm}

In order to properly marginalize constraints over cosmological parameters, 
we need a prediction of the mass function
enhancement as a function of cosmological parameters and the
$f(R)$ parameter $f_{R0}$. Due to the large computing requirements for
full $f(R)$ simulations \cite{Oyaizu:2008sr}, running simulations for a range
of parameters is not feasible, and we use a model of the mass function
enhancement based on spherical collapse and the Sheth-Tormen (ST)
prescription (see \cite{halopaper} for details).  Note that we use this prescription and
the cosmological simulations that calibrate it to
compute enhancements only.   For the $\Lambda$CDM baseline predictions,
we use  mass function results from state-of-the-art numerical simulations
 (see \S \ref{sec:clusterlikelcdm}). 

The ST description 
 for the comoving number density of halos per logarithmic interval in the virial mass $M_{\rm v}$ is given by
\begin{align}
n_{\ln M_{\rm v}} \equiv
\frac{d n}{d\ln M_{\rm v}} &= {\bar \rho_{\rm m} \over M_{\rm v}} f(\nu) {d\nu \over d\ln M_{\rm v}}\,, 
         \label{eqn:massfn}
\end{align}
where the peak threshold $\nu = \delta_c/\sigma(M_{\rm v})$ and 
\begin{eqnarray}
\nu f(\nu) = A\sqrt{{2 \over \pi} a\nu^2 } [1+(a\nu^2)^{-p}] \exp[-a\nu^2/2]\,.
\end{eqnarray}
Here
$\sigma(M)$ is the variance of the linear density field convolved with a top hat of radius $r$
that encloses $M=4\pi r^3 \bar \rho_{\rm m}/3$ at the background density
\begin{eqnarray}
\sigma^2(r) = \int \frac{d^3k}{(2\pi)^3} |\tilde{W}(kr)|^2 P_{\rm L}(k)\,,
\label{eq:sigmaR}
\end{eqnarray}
where $P_{\rm L}(k)$ is the linear power spectrum and $\tilde W$ is the Fourier transform
of the top hat window.  The normalization constant $A$ is chosen 
such that $\int d\nu f(\nu)=1$. The parameter values of $p=0.3$, $a=0.75$, and
$\delta_c=1.673$ for the spherical collapse threshold have previously been shown to 
match simulations of $\Lambda$CDM at the $10-20\%$ level. 
The virial mass is defined as the mass enclosed at 
the virial radius $r_{\rm v}$, at which the average density is $\Delta_{\rm v}$
times the critical density $\rho_{\rm cr}$. For consistency with cluster analyses, overdensities
will refer to critical density throughout the paper; the corresponding
overdensities in terms of the background matter density are given by
$\Delta_{\rho_m} = \Delta_{\rho_{\rm cr}} / \Om$.

\begin{figure}[t!]
\centering
\includegraphics[width=0.48\textwidth]{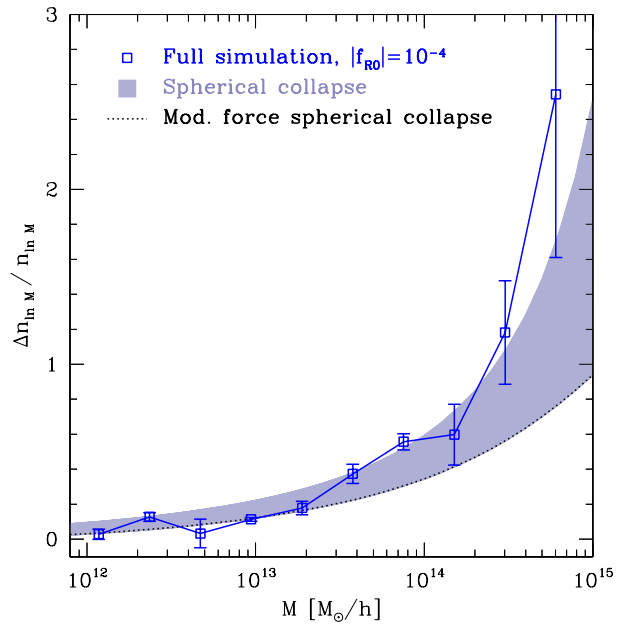}
\caption{{\small Mass function enhancement at $z=0$ with respect to $\Lambda$CDM 
as a function of $M=M_{500}$, measured in $f(R)$
simulations with $|f_{R0}|=10^{-4}$. Also shown is the range of spherical 
collapse predictions from \cite{halopaper}.
For the constraints, we conservatively use the lower limit of the shaded band 
(dashed line).}
\label{fig:dndmSim}}
\end{figure}

Ref.~\cite{halopaper} derived a model for the mass function enhancement
measured in the $f(R)$ N-body simulations. The mass function calculation
is based on the Sheth-Tormen prescription using the {\it linear} power
spectrum for the $f(R)$ model, and two limiting cases for the spherical
collapse parameters. In one case, we simply assume that the
spherical perturbation considered is always larger than the (local)
Compton wavelength of the $f_R$ field, so that gravity is GR throughout, 
and the spherical
collapse parameters are unchanged ($\delta_c=1.673$ and $\Delta_{\rm v}=94$
for  $\Om=0.24$). In the second case, we assume that the perturbation is
always smaller than the local Compton wavelength, so that forces are
enhanced by 4/3. The corresponding
modified spherical collapse parameters are $\delta_c=1.692$ and $\Delta_{\rm v}=74$
for  $\Om=0.24$. \reffig{dndmSim} shows the range of predicted mass function
enhancement for these two limiting cases, and the results of the 
$f(R)$ simulations, for $|f_{R0}|=10^{-4}$. The mass definition used,
$M \equiv M_{\Delta}$ with $\Delta=500$ relative to the critical density, is
the same as used in the X-ray cluster measurements. In order to obtain
conservative upper limits, we choose the modified force prediction
for the mass function, which corresponds to the lower bound of the
shaded band in \reffig{dndmSim}.

For a given set of cosmological parameters ($A_s$, $\Om$, $h$, $f_{R0}$),
we first calculate the Sheth-Tormen mass functions $dn/d\ln M_{\rm v}$ 
for $\Lambda$CDM and $f(R)$ using the respective linear power spectra.
We then rescale each mass function from the respective virial mass to
the common mass definition $M=M_{500}$, using the procedure outlined in
\cite{HuKravtsov}. We need to assume a halo profile for this mass
rescaling, which we take to be of the NFW form with the concentration
relation given by \cite{BullockEtal}. As shown in \cite{halopaper}, the 
profiles of dark matter halos in $f(R)$ within $r_{\rm v}$ are sufficiently similar
to those measured in GR simulations that $f(R)$ effects can be neglected in the mass
rescaling. Finally, $\Delta_{\rm v}$ for $\Lambda$CDM is obtained from
the fitting formula of \cite{EkeEtal}:
\be
\Delta^{\rm GR}_{\rm v}(\Om) = 178\; \Om^{0.45}.
\ee 
For the $f(R)$ enhanced forces, we assume the same scaling, fixing
the ratio $\Delta^{f(R)}_{\rm v}/\Delta^{\rm GR}_{\rm v} = 74/94$.
We neglect the small $\Om$ dependence of the collapse thresholds
within the range of interest, and keep $\delta_c(\Lambda\rm CDM)=1.673$
and $\delta_c(f(R)) = 1.692$ fixed.

\section{Cluster Likelihood}
\label{sec:clusters}

In this section we describe the cluster likelihood as a function of cosmological
and $f(R)$ parameters.
Since we assume a spatially flat cosmology and the expansion history of 
$f(R)$ models are indistinguishable from $\Lambda$CDM,
the main cosmological parameters that we have to consider are
$\Om$, $h$, and the primordial normalization $A_s$ at $k=0.02$ Mpc$^{-1}$. Other
parameters such as the power spectrum tilt do not affect the constraints
appreciably \cite{VikhlininIII}. Since the CMB determines $\Om h^2$ to good 
precision, we are mainly dealing with
$\Om$, $A_s$, and the $f(R)$ parameter $f_{R0}$ for the cluster abundance.

In \S \ref{sec:clusterlikelcdm}, we review the likelihood approach taken in
Ref.~\cite{VikhlininIII}.  In \S \ref{sec:clusterlikefr}, we describe the rescaling of
these results for $f(R)$ models.  

\subsection{$\Lambda$CDM}
\label{sec:clusterlikelcdm}

The cluster sample used in this work is the low-$z$ subsample of 49
clusters described in \cite{VikhlininII,VikhlininIII}. This is an
X-ray flux-limited sample of clusters originally detected in the
\emph{ROSAT} All-Sky Survey at high Galactic latitudes and at
$z>0.025$. All of the objects were later observed with \emph{Chandra},
providing low-scatter proxies for the total mass which can be
constructed from the mean X-ray temperature and gas mass (see
below). The effective redshift depth of this sample is $z<0.15$.

Cluster total masses are estimated using the $Y_{X}$ parameter defined
as $Y_{X}=T_{X}\times M_{\text{gas}}$, where $T_{X}$ is the average
temperature measured from the integral X-ray spectrum, and
$M_{\text{gas}}$ is the estimated gas mass derived from the analysis
of the X-ray surface brightness profile. $Y_{X}$ is a direct X-ray
observable, even though it is hard to backtrack it to raw observables
such as the total X-ray luminosity. For a detailed description of the
data analysis procedures, see \cite{VikhlininII}.

Based on state-of-the-art cosmological numerical simulations, $Y_{X}$
is expected to be tightly correlated with the total cluster mass, $M
\propto Y_{X}^{3/5}\,H(z)^{-2/5}$, with $<10\%$ scatter
\cite{2006ApJ...650..128K}. Numerical experiments show that the
$Y_{X}-M$ relation is remarkably insensitive to the cluster dynamical
state. The power law slope and evolution factor are also insensitive
to the details of star formation history and non-gravitational heating
of the intracluster medium although these processes do change the
overall normalization of the relation (e.g,
\cite{2006ApJ...650..128K}, \cite{2009ApJ...700..989B}). The
normalization of the $Y_{X}-M$ relation was determined observationally
\cite{VikhlininII} using two independent techniques, (1) by X-ray
hydrostatic method using a subsample of dynamically relaxed clusters,
and (2) by weak lensing mass measurements. The two methods are in
excellent agreement, and this was used to place an upper limit on
systematic uncertainties in the cluster mass scale, $\Delta
M/M<9\%$. For our purposes, this agreement also means that the
normalization of the $Y_{X}-M$ relation is tied to the weak lensing
measurements, which should provide the true mass in the $f(R)$ theories we
consider  (\S\,\ref{sec:model}).

The cluster component of the likelihood function is computed assuming
a purely Poissonian nature of statistical fluctuations\footnote{We
  ignore a contribution from cosmic variance; the validity of this
  assumption is justified in \cite{VikhlininII}.}:
\begin{equation}\label{eq:likelihood}
  \begin{split} 
  \ln L = \sum_i \ln\, & p(\Mest_i,z_i) + \sum_i \ln \Mest_i\; \\
  & -\int dz\int d\Mest  p(\Mest,z) \,,
  \end{split}
\end{equation}
where $\Mest_i$ and $z_{i}$ is the estimated mass and redshift of
cluster $i$, $p(\Mest_i,z_i)$ is the model probability density to observe a
cluster with mass $\Mest_{i}$ at redshift $z_{i}$, and summation is
over the clusters in the sample and integration is over pre-selected
$\zmin-\zmax$ and $\Mmin-\Mmax$ intervals. 
Note that the $\ln \Mest_i$ term appears because the mass estimates
and hence the mass binning required to convert probability densities into 
probabilities depends on cosmology.

The model probability density, $p$,
is a product of the Tinker et al.\ mass function model
\cite{TinkerEtal} for a given set of $\LCDM$ parameters, cosmological
$dV/dz$ function, and the survey selection probability as a function
of mass and redshift. The product of these components is convolved
with the intrinsic and measurement scatters in the
$Y_{X}-M$ relation. The computation of all these terms is
  discussed in detail in \cite{VikhlininII}, and the $\LCDM$ parameter
  constraints derived from this cluster dataset are presented in
  \cite{VikhlininIII}. We now turn to the computation of the cluster
  likelihood in the $f(R)$ models.

\begin{figure}[t!]
\centering
\includegraphics[width=0.48\textwidth]{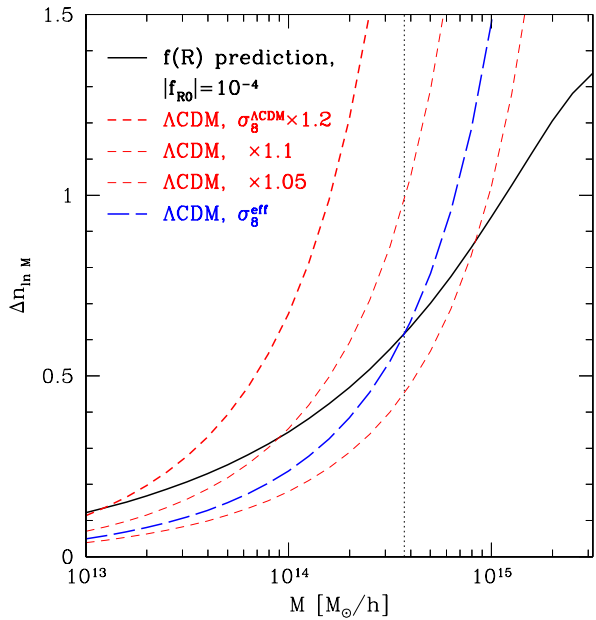}
\caption{{\small Mass function enhancement for \fRzero from the spherical
collapse model (black, solid) as in \reffig{dndmSim}, and the corresponding
enhancement when increasing the linear power spectrum normalization
in $\Lambda$CDM. The vertical line indicates the pivot mass $\Mxeff$ used
to calculate the likelihood. The blue dashed line shows the enhancement for
a rescaled normalization ($\seff=\sigma_8 \times 1.066$) 
that matches the $f(R)$ enhancement at $\Mxeff$.}
\label{fig:dndm}}
\end{figure}

\begin{figure}[t!]
\centering
\includegraphics[width=0.48\textwidth]{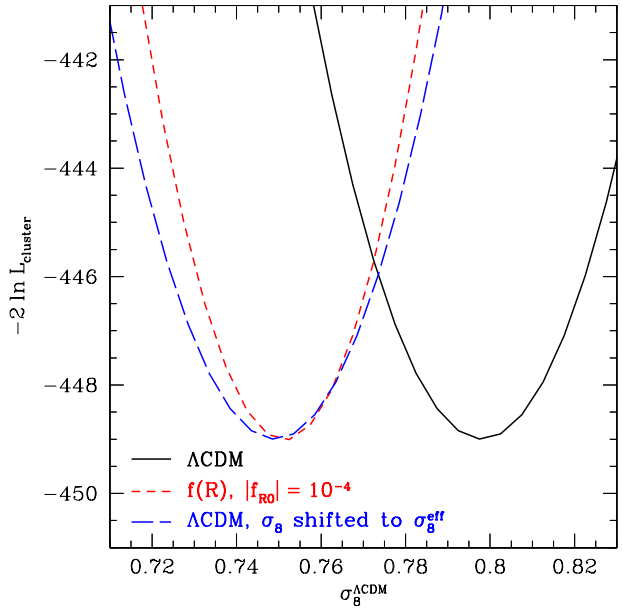}
\caption{{\small Cluster likelihood as function of primordial normalization
(quantified by the linear power spectrum normalization $\sigma_8^{\LCDM}$ a
$\Lambda$CDM model would give), for fixed values of
$\Om=0.258$, $h=0.716$, and \fRzero in case of the $f(R)$ prediction.
The red short-dashed line shows the likelihood calculated using the full $f(R)$ mass function
enhancement, while the blue long-dashed line shows the $\Lambda$CDM likelihood obtained 
with the rescaled normalization, $\seff$.}
\label{fig:clusterL}}
\end{figure}

\subsection{$f(R)$ Scaling}
\label{sec:clusterlikefr}

A full likelihood analysis of the $f(R)$ cluster constraints would
entail a recalculation of the X-ray cluster likelihood grid
\cite{VikhlininIII}. Since the modification to the {\it shape} of the
cluster mass function in $f(R)$ is not dramatic across the dynamic
range of masses probed by the cluster mass function data (e.g.,
Fig.16,~17 in~\cite{VikhlininII}), we opt for a simpler approach: for
each point in parameter space $\Om,A_s,f_{R0}$, we calculate the
$f(R)$ mass function enhancement at a pivot mass, $\Mxeff =
3.677\times 10^{-14}\:\Msunh$.  Here we take $M$ to be the true or
lensing mass, which is the most conservative assumption.  
Equating the dynamical mass $M_{\rm
  dyn}\approx 4M/3$ to the $Y_X$ based mass can only increase the
abundance enhancement in the $f(R)$ models.  We correspondingly also
ignore the additional tension in $f(R)$ implied by the observed agreement 
between the lensing and $Y_X$ masses.

Then, for this set of
parameters, the enhancement is converted to an {\it effective}, not
actual, linear power spectrum normalization, $\seff$, assuming a
$\Lambda$CDM model with the same $\Om$ and the mass function
prescription of \cite{TinkerEtal}. This approximation assumes that, in
the mass range probed by the X-ray clusters, the mass-dependence of
the mass function enhancement due to $f(R)$ has the same shape as that
due to an increased power spectrum normalization. This is only
approximately true (see \reffig{dndm}), since the growth is
scale-dependent in $f(R)$.

In order to benchmark the accuracy of this simplified approach,
we recalculated the cluster likelihood for fixed values of $\Om,h$
and a range of $A_s$ using the full $f(R)$ mass function, for
$|f_{R0}|=10^{-4}$. We then
compared this likelihood with the $\Lambda$CDM cluster likelihood 
calculated for the rescaled normalization, $\seff$. The resulting
likelihoods are shown as function of the primordial normalization
in \reffig{clusterL}. First, clusters clearly prefer a lower
primordial normalization in the $f(R)$ model, due to the enhanced growth.
The approximate likelihood calculated using the rescaled normalization
agrees quite well with the full likelihood. Note that this approach
is in any case conservative, as the constraints are weakened (though
not significantly) by neglecting the additional information.

While all $f(R)$ mass function enhancements were calculated at $z=0$, we
verified that the evolution of enhancements in the redshift range
probed by the cluster sample, $z=0-0.15$, is negligible.

\reffig{clusterL} shows that if $\Om \approx 0.26$ and the primordial normalization
of the simulations  
were verified to very high precision by external constraints, and the 
mass calibration of
the cluster sample carried no systematic error, the cluster
abundance would be able to rule out the $f(R)$ model with \fRzero
at around 95\% confidence. We next address to what extent these expectations
apply to the joint cosmological and cluster data.

\section{Combined Constraints}
\label{sec:combined}

The excess cluster abundance predicted in $f(R)$ models can be converted into
limits on the field amplitude  $f_{R0}$ once data external to the clusters
has fixed the background expansion history and primordial normalization of density 
fluctuations.   In \S \ref{sec:external} we describe the external data that
we use for these purposes and present combined results in \S \ref{sec:results}.

\subsection{External Data Sets}
\label{sec:external}

{\it CMB:} Following Ref.~\cite{VikhlininIII}, we employ a simplified approach to incorporating
CMB constraints from WMAP5 into the cluster analysis.
We take
three CMB parameters --- angular scale of the first acoustic peak, $\ell_A$;
the so called shift parameter, $R$; and the recombination redshift, $z_*$.
The likelihood for the geometric side of the WMAP-5 data is computed using the covariance
matrix for $\ell_A$, $R$, and $z_*$ provided in
\cite{Kometal08}.  In the $\Lambda$CDM expansion history these
quantities depend on $\Omega_m$, $h$ and $\Omega_b h^2$.  
 
Next we add the information contained on the initial amplitude of fluctuations.   
 The WMAP team provides the
amplitude of the curvature perturbations at the $k=0.02$~Mpc$^{-1}$
scale,
\begin{equation}
  \hat A_s = (2.21\pm0.09)\times10^{-9} \,.
\end{equation}
To implement this constraint in terms of $\sigma_8^{\Lambda{\rm CDM}}$ and
the chosen cosmological parameters we use the fitting formula
\cite{Hu:2003pt}:
\begin{equation}\label{eq:CMB:s8:fit}
  \begin{split}
    A_s^{{1/2}} \approx & \frac{\sigma_8^{\Lambda{\rm CDM}}}{1.79\times10^{4}}\,
    \left(\frac{\Omega_b h^2}{0.024}\right)^{1/3}
    \left(\frac{\Omega_m h^2}{0.14}\right)^{-0.563} \\
    & \times\;(7.808\,h)^{(1-n)/2}\,
    \left(\frac{h}{0.72}\right)^{-0.693}
    \frac{0.76}{G_0}
  \end{split}
\end{equation}
(we adjusted numerical coefficients to take into account that the
original fit uses the CMB amplitude at
$k=0.05$~Mpc$^{-1}$ while the WMAP-5 results are reported for
$k=0.02$~Mpc$^{-1}$). In this equation, $G_0$ is the growth suppression relative to
$\delta \propto (1+z)^{-1}$ due
to $\Lambda$ evaluated today.
We then include a $\chi^2$ contribution given
by
\begin{equation}
  \chi^2_{\rm CMBnorm} = (A_s \times 10^9 -
  2.21)^2/0.09^2.
\end{equation}
The $\chi^2_{\rm CMBnorm}$ component is computed assuming a fixed
$n=0.95$; the results are completely insensitive to variations of $n$
within the WMAP measurement uncertainties and even to setting
$n=1$. The sum of the geometric and growth component of the CMB
$\chi^{2}$ is marginalized over $\Omega_b h^2$ and $h$. The end result
is a $\chi^2_{\rm CMB}$ for the CMB that is a function of $\Omega_m$
and $\sigma_8^{\Lambda{\rm CDM}}$.

\begin{figure}[t]
\centering
\includegraphics[width=0.48\textwidth]{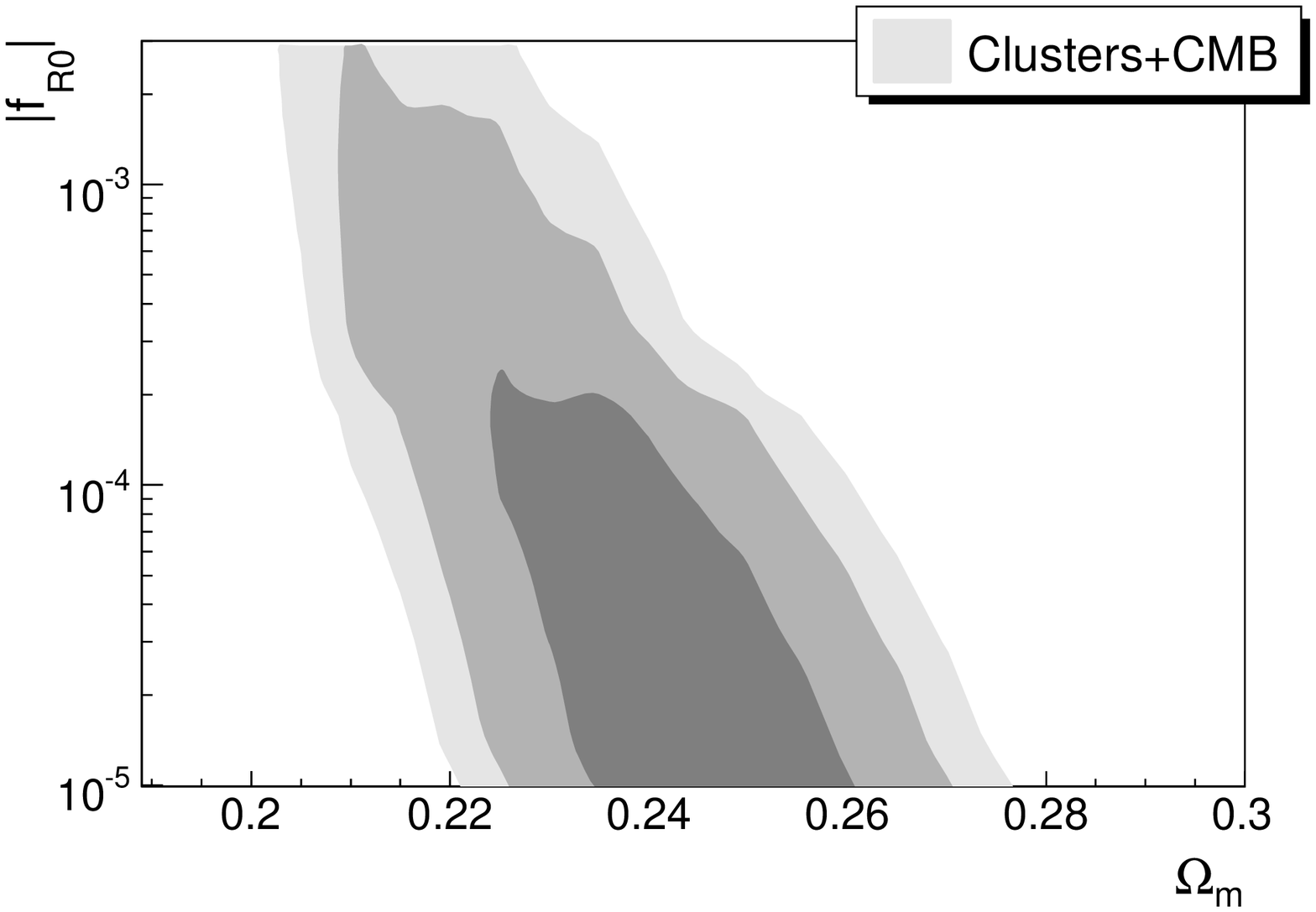}
\includegraphics[width=0.48\textwidth]{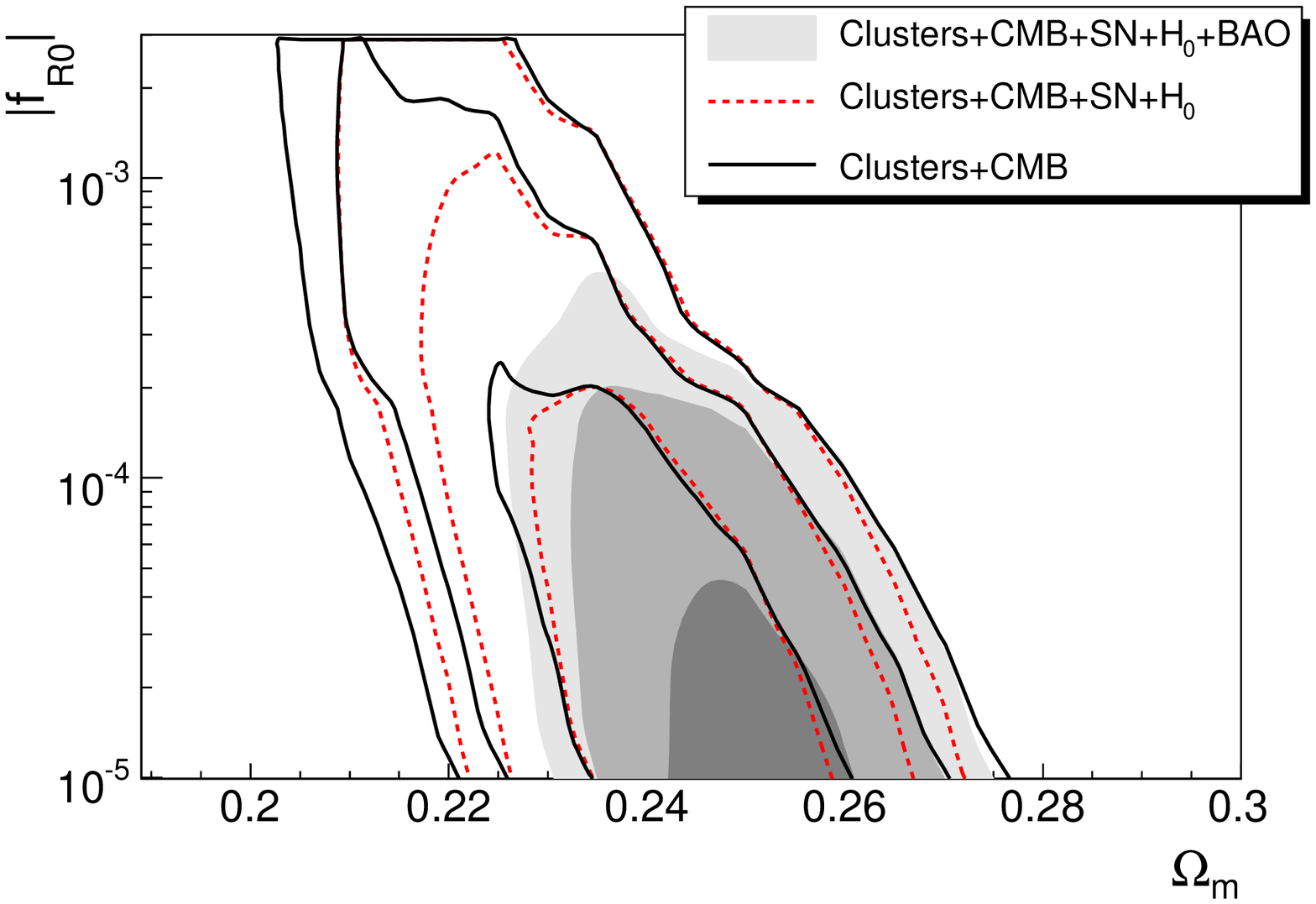}
\caption{{\small \textit{Top panel:} Likelihood contours from clusters+CMB in the 
$f_{R0}-\Om$ plane, marginalized over the primordial normalization.
Shown are 68.3\%, 95.4\% and 99.7\% likelihood contours.
\textit{Bottom panel:} Likelihood contours in the $f_{R0}-\Om$ plane marginalized
over the primordial normalization, for clusters+CMB only and when including
other geometric measurements.}
\label{fig:fR-Om}}
\end{figure}

{\it SN:}
We use the distance moduli estimated for the Type~Ia
supernovae from the HST sample \cite{Riess:2006fw}, SNLS
survey \cite{Astier:2005qq}, and ESSENCE survey
\cite{WoodVasey:2007jb}, combined with the nearby supernova sample
as compiled in Ref.~\cite{Davis:2007na}. Calculation of the SN~Ia component of the
likelihood function for the given cosmological model is standard and can
be found in any of the above references.
The end result is a $\chi^2_{\rm SN}$ that depends on $\Omega_m$ only.

{\it $H_0$:}
We use the recent determination of $H_0$ \cite{RiessEtal09},
$H_0=74.2\pm4.8\:\rm km/s/Mpc$, in conjunction with the CMB constraint
of $\Om h^2 = 0.133 \pm 0.006$ \cite{Komatsu09}
as a measurement of $\Om$. Marginalizing
over the uncertainty in $\Om h^2$ results in the 
following Gaussian likelihood:
\begin{equation}
\chi^2_{H_0} = \left (\frac{\Om - 0.242}{0.034}\right )^2.
\end{equation}

{\it BAO:}
In a similar way, we use the distance scale given by the baryon acoustic
oscillation measurements (BAO) of \cite{PercivalEtal09}. We use their
Eq.~(16), which after marginalizing over $\Om h^2$ yields:
\begin{equation}
\chi^2_{\rm BAO} = \left (\frac{\Om - 0.285}{0.019}\right )^2.
\end{equation}
The BAO constraint is in fact the most precise one and hence dominates
our $\Om$ likelihood. 

Finally, we combine all the contributions of external data sets
\begin{equation}
\chi^2_{\rm ext} = \chi^2_{\rm CMB} + \chi^2_{H_0} + \chi^2_{\rm SN} + \chi^2_{\rm BAO},
\end{equation}
and add $\chi^2_{\rm ext}$ to the cluster likelihood contribution of Eq.~(\ref{eq:likelihood}),
$\chi^2_{\rm cl} \equiv -2\ln L$.

\subsection{Results}
\label{sec:results}

In Fig.~\ref{fig:fR-Om} (top) we show the results of combining the cluster abundance
and CMB constraints.  The assumption of spatial flatness in combination with the
CMB data alone constrains $\Om$ and limits the extent of the $f_{R0}-\Om$ degeneracy.
Note that the bounds on $f_{R0}$ tighten as $\Om$ increases. 
With only these two data sets the statistical upper limit after marginalizing 
over $\Om$ is $|f_{R0}|/10^{-4} < 8$ at 95\% statistical confidence level (CL).

\begin{figure}[t]
\centering
\includegraphics[width=0.48\textwidth]{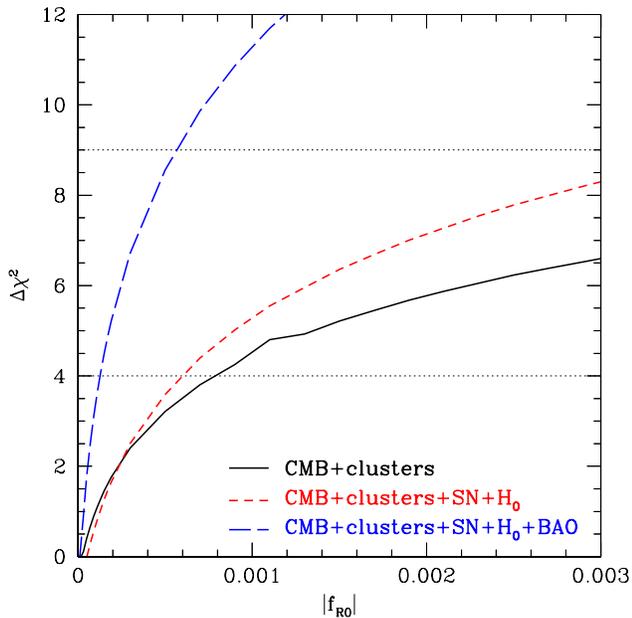}
\caption{{\small Likelihood relative to $f_{R0}=0$ ($\Lambda$CDM) as a 
function of $f_{R0}$ for CMB+clusters and 
including other geometric measures. We have marginalized over $\Om$ and
the primordial normalization. The horizontal lines show the $2\sigma$ and
$3\sigma$ confidence levels. Using all measures combined, 
$|f_{R0}|/10^{-4} <1.3$ at 95\% confidence level.}
\label{fig:delta-chi2}}
\end{figure}

Data on SN distances, $H_0$ and BAO distances tighten the bounds on $\Om$ 
reducing the degeneracy with $|f_{R0}|$.  
Fig.~\ref{fig:fR-Om} (bottom) shows that the BAO data in particular make a strong impact on
constraints since they favor high $\Om$.   In Fig.~\ref{fig:delta-chi2} we show the $\delta\chi^2$ statistic after marginalization of $\Om$.  With all of the data, $|f_{R0}|/10^{-4} < 1.3$ at 95\% statistical CL.
Table~\ref{tab:res} summarizes the upper limits on $f_{R0}$ for the different
data sets and for different confidence levels.

\begin{figure}[t!]
\centering
\includegraphics[width=0.48\textwidth]{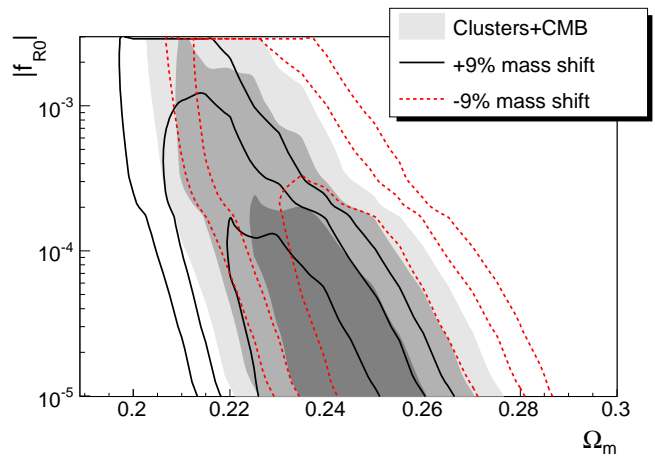}
\caption{{\small Likelihood contours from clusters+CMB in the 
$f_{R0}-\Om$ plane, marginalized over the primordial normalization
as shown in \reffig{fR-Om}.
The solid contours show the results for the standard cluster mass scale,
while the black solid (red dashed) lines show the likelihood in case cluster 
masses are overestimated (underestimated) by 9\%.}
\label{fig:fR-Om-syst}}
\end{figure}

The main caveat to these statistical constraints is the possibility of systematic shifts in
the mass calibration of the cluster sample. 
In Fig.~\ref{fig:fR-Om-syst} we show the
impact of $\pm 9\%$ shifts in the cluster mass scale on the cluster+CMB 
constraints. Note that this effect mainly shifts the contours by 
$\Delta \Om \sim \pm 0.015$. If we assume that cluster masses are
{\it underestimated} (``$-9\%$ mass shift'', i.e. $M_{X,\rm obs}=0.91\: M$),
the abundance at a fixed mass is in fact higher, and hence allows higher 
$f_{R0}$ values. Conversely, in the case that cluster masses are 
{\it overestimated} (``+9\% mass shift'', i.e. $M_{X,\rm obs}=1.09\: M$), the true 
abundance at a fixed mass 
is smaller, hence tightening $f_{R0}$ constraints.   

\begin{figure}[t]
\centering
\includegraphics[width=0.48\textwidth]{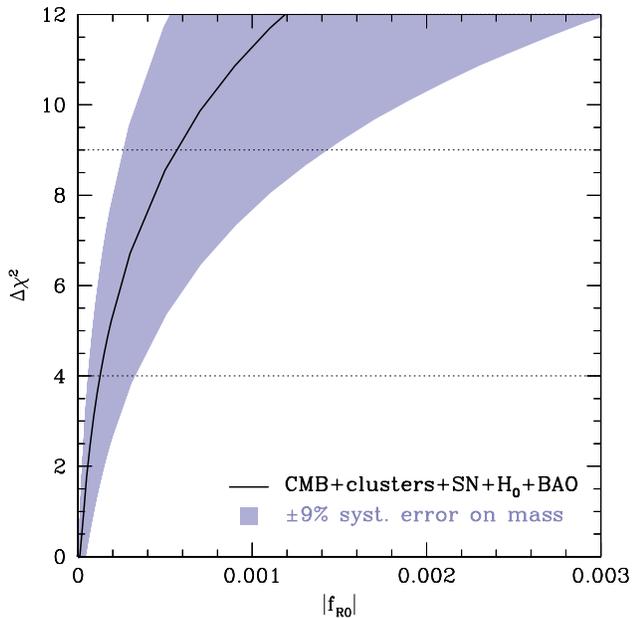}
\caption{{\small Likelihood as a function of $f_{R0}$ as in \reffig{delta-chi2}
for CMB+clusters and all geometric measures. The shaded band shows
the weakening/strengthening of the constraints when varying the absolute
mass scale by $\pm 9$\%, corresponding to the estimated systematic 
uncertainty. In the weakened case (masses underestimated), the constraint
degrades to $|f_{R0}|/10^{-4} < 3.0$ at 95\% confidence level.}
\label{fig:delta-chi2-syst}}
\end{figure}

We show the impact of a $\pm 9\%$ mass calibration error on the final joint results
in Fig.~\ref{fig:delta-chi2-syst}, and in Tab.~\ref{tab:res}.   
The net result is that the 95\% statistical CL carries
systematic errors of $+1.7/-0.6\times 10^{-4}$, which we shall write as
$|f_{R0}|/10^{-4} < 1.3^{+1.7}_{-0.6}$.

Note that given $M_{\rm dyn} \approx
4M/3$ in $f(R)$ if there is no screening due to the chameleon mechanism, 
the X-ray measurements possibly {\it overestimate} cluster masses by up to
$33\%$. Hence, we expect that our use of lensing masses in calibrating the 
enhancement makes our constraint conservative
even given the possibility of a $9\%$ underestimate in the X-ray mass measurement.

Our model of the mass function enhancements (\S \ref{sec:dndm})
also represents a lower bound which always underpredicts the enhancement measured
in N-body simulations for $10^{-6}< |f_{R0}| < 10^{-3}$ \cite{halopaper}.
We have also determined upper limits on $|f_{R0}|$ using
the less conservative limiting case presented in \cite{halopaper},
which corresponds to using alternate collapse parameters that correspond to the GR
values of $\delta_c$ and $\Delta_v$.
This case is shown as the upper boundary of the shaded band in \reffig{dndm}.
While the prediction is still below the simulations, and a better fit,
for $|f_{R0}|\gtrsim 10^{-4}$, it overpredicts the enhancements for smaller
field values \cite{halopaper}. The last line in \reftab{res} shows the
resulting limits, which are tighter by a factor of 3--4.
We cannot guarantee that this  model does not
overpredict the enhancement in some region of the parameter space involved,
but the corresponding tightening of the constraints again indicates that our quoted limit
should be viewed as conservative.

\begin{table}[b!]
\caption{Upper limits on $f_{R0}$ in units of $10^{-4}$.\label{tab:res}} 
\begin{center}
  \leavevmode
  \begin{tabular}{l|c|c|c}
\hline
Confidence level (CL) & 68.3\% & 95.4\% & 99.7\% \\
\hline\hline
Clusters+CMB & 1.0 & 7.9 & $> 31$ \\
Clusters+CMB+SN+$H_0$ & 1.0 & 5.4 & $> 31$ \\
\textbf{Clusters+CMB+SN+$H_0$+BAO}\  \  & {\bf 0.3} & {\bf 1.3} & {\bf 6.3} \\
\hline
\quad with $+9\%$ mass shift & 0.2 & 0.7 & 3.1 \\
\quad{with $-9\%$ mass shift} & {0.7} & {3.0} & {14.7} \\
\quad{with alternate collapse parameters} & 0.09 & 0.4 & 1.8 \\
\hline
\end{tabular}
\end{center}
\end{table}

\section{Discussion}
\label{sec:discussion}

We have provided the first, simulation calibrated, cluster abundance 
constraints on a modified gravity model, specifically  
$f(R)$ gravity.
Enhanced forces below the Compton wavelength of the scalar
field lead to corresponding enhancements in the cluster abundance, making
the latter a sensitive probe of gravity on cosmological scales.

Combined with constraints on the primordial amplitude of fluctuations from the CMB,
and geometric constraints from  CMB, supernovae, $H_0$ and BAO, the cluster abundance
provides the most stringent constraint on these enhancements to date.    
In terms of the field amplitude today, the constraint is
$|f_{R0}|/10^{-4} < 1.3^{+1.7}_{-0.6}$, 
where the range reflects a $\mp 9\%$ mass calibration error, an
improvement over previous constraints by 4 orders of magnitude.    
This corresponds to an upper limit on the range of the gravitational
force modifications of $\lambda_C < 38^{+19}_{-10}$~Mpc in this $f(R)$ model.

Our constraint should be viewed as conservative even given the $9\%$ mass calibration 
error.   We have ignored an overestimate  of dynamically based $X$-ray masses over true or
lensing masses, which could be up to a $33\%$ shift that would further enhance the abundance and strengthen the constraint.
In addition, we have not considered the possibility of constraining 
$f(R)$ force modifications from the comparison of cluster lensing and
dynamical masses.

Furthermore, our model of the mass function enhancements 
represents a conservative lower bound which always underpredicts the 
enhancement measured in N-body simulations. We found that the constraints
tighten significantly if we use a less conservative model but a robust implementation
would require more accurate mass function calibration across the parameter space.

On the observational side,
current constraints are limited mainly by systematics in the mass calibration and secondarily
by the small number of local clusters.
Relaxing the assumption of a flat universe is not expected to degrade the
upper limits appreciably. This is because our constraints only depend strongly
on the allowed range in $\Om$ and marginalizing over curvature
changes this range negligibly once BAO are combined with SN and/or the CMB
 \cite{PercivalEtal09}.

In the future, the abundance of massive clusters could ultimately provide 
another order of magnitude
improvement in the field amplitude to $|f_{R0}| \sim 10^{-5}$, rivaling solar system
tests of gravity but in a very different, low curvature regime \cite{HuSaw07a}.   
Below field values of $\sim 10^{-5}$,
the chameleon mechanism suppresses the enhancement at the high mass end
\cite{halopaper}. 
In this regime, further
improvements are potentially available if the abundance of galaxy groups can provide
constraints on the halo abundance at intermediate masses.

While we have considered a specific functional form of $f(R)$ here \cite{HuSaw07a},
different functional forms have been proposed in the literature, see e.g.
\cite{ApplebyBattye,Starobinsky07,NojOdin07}. 
These models differ primarily in the evolution with redshift of the
Compton wavelength of the $f_R$ field. Hence, we expect our
 results to be generic once the field amplitude and range are rescaled to some
 effective redshift which matches the impact on the linear growth today on a scale
 relevant for clusters.   For example in models where the curvature dependence
 is steeper,  the field amplitude today is allowed to be larger since its current value
 has little impact on the growth of structure.   In these models solar system tests
 become even more powerful relative to cosmological tests.

More generally, the abundance of galaxy clusters promises
to be a good probe of other modified gravity scenarios as well, such as DGP and other 
braneworld models once their mass functions are calibrated by cosmological
simulations \cite{KW,Schmidt09}.

 \smallskip
 \noindent {\it Acknowledgments}:  We thank the Aspen Center for Physics where part of this
 work was completed. FS and WH were supported by
 the Kavli Institute for Cosmological Physics (KICP) at the University
 of Chicago through grants NSF PHY-0114422 and NSF PHY-0551142 and an
 endowment from the Kavli Foundation and its founder Fred Kavli.  WH
 was additionally supported by U.S.~Dept.\ of Energy contract
 DE-FG02-90ER-40560 and the David and Lucile Packard Foundation. AV
 was supported by NASA grants and contracts NAG5-9217, GO5-6120A,
 GO6-7119X and NAS8-39073.
Computational resources for the cosmological simulations were provided by the KICP-Fermilab
computer cluster.

\vfill
\bibliographystyle{arxiv_physrev}
\bibliography{frcluster}

\def\eprinttmppp@#1arXiv:@{#1}
\providecommand{\arxivlink[1]}{\href{http://arxiv.org/abs/#1}{arXiv:#1}}
\providecommand{\arxivlinknopre[1]}{\href{http://arxiv.org/abs/#1}{#1}}
\providecommand{\eprintmod}[1][XXXX.XXXX]{\IfSubStr{#1}{arXiv}{\arxivlinknopre%
{#1}}{\arxivlink{#1}}}
\providecommand{\adsurl}[1]{\href{#1}{ADS}}
\begin{thebibliography}{33}
\expandafter\ifx\csname natexlab\endcsname\relax\def\natexlab#1{#1}\fi
\expandafter\ifx\csname bibnamefont\endcsname\relax
  \def\bibnamefont#1{#1}\fi
\expandafter\ifx\csname bibfnamefont\endcsname\relax
  \def\bibfnamefont#1{#1}\fi
\expandafter\ifx\csname citenamefont\endcsname\relax
  \def\citenamefont#1{#1}\fi
\expandafter\ifx\csname url\endcsname\relax
  \def\url#1{\texttt{#1}}\fi
\expandafter\ifx\csname urlprefix\endcsname\relax\def\urlprefix{URL }\fi

\bibitem{Caretal03}
S.~M. Carroll, V.~Duvvuri, M.~Trodden and M.~S. Turner,
\newblock Phys. Rev. {\bf D70}, 043528 (2004), [\eprintmod[astro-ph/0306438]].

\bibitem{NojOdi03}
S.~Nojiri and S.~D. Odintsov,
\newblock Phys. Rev. {\bf D68}, 123512 (2003), [\eprintmod[hep-th/0307288]].

\bibitem{Capozziello:2003tk}
S.~Capozziello, S.~Carloni and A.~Troisi,
\newblock Recent Res. Dev. Astron. Astrophys. {\bf 1}, 625 (2003),
  [\eprintmod[astro-ph/0303041]].

\bibitem{khoury04a}
J.~{Khoury} and A.~{Weltman},
\newblock \prd {\bf 69}, 044026 (2004), [\eprintmod[astro-ph/0309411]].

\bibitem{HuSaw07a}
W.~{Hu} and I.~{Sawicki},
\newblock \prd {\bf 76}, 064004 (2007), [\eprintmod[0705.1158]].

\bibitem{SonPeiHu07}
Y.-S. {Song}, H.~{Peiris} and W.~{Hu},
\newblock \prd {\bf 76}, 063517 (2007), [\eprintmod[0706.2399]].

\bibitem{halopaper}
F.~{Schmidt}, M.~{Lima}, H.~{Oyaizu} and W.~{Hu},
\newblock \prd {\bf 79}, 083518 (2009), [\eprintmod[0812.0545]].

\bibitem{BekSan94}
J.~D. {Bekenstein} and R.~H. {Sanders},
\newblock \apj {\bf 429}, 480 (1994), [\eprintmod[astro-ph/9311062]].

\bibitem{Oyaizu:2008sr}
H.~Oyaizu,
\newblock Phys. Rev. {\bf D78}, 123523 (2008), [\eprintmod[arXiv:0807.2449]].

\bibitem{Oyaizu:2008tb}
H.~Oyaizu, M.~Lima and W.~Hu,
\newblock Phys. Rev. {\bf D78}, 123524 (2008), [\eprintmod[0807.2462]].

\bibitem{HuKravtsov}
W.~{Hu} and A.~V. {Kravtsov},
\newblock \apj {\bf 584}, 702 (2003), [\eprintmod[astro-ph/0203169]].

\bibitem{BullockEtal}
J.~S. {Bullock} {\em et~al.},
\newblock \mnras {\bf 321}, 559 (2001), [\eprintmod[astro-ph/9908159]].

\bibitem{EkeEtal}
V.~R. {Eke}, J.~F. {Navarro} and C.~S. {Frenk},
\newblock \apj {\bf 503}, 569 (1998), [\eprintmod[astro-ph/9708070]].

\bibitem{VikhlininIII}
A.~{Vikhlinin} {\em et~al.},
\newblock \apj {\bf 692}, 1060 (2009), [\eprintmod[0812.2720]].

\bibitem{VikhlininII}
A.~{Vikhlinin} {\em et~al.},
\newblock \apj {\bf 692}, 1033 (2009), [\eprintmod[0805.2207]].

\bibitem{2006ApJ...650..128K}
A.~V. {Kravtsov}, A.~{Vikhlinin} and D.~{Nagai},
\newblock \apj {\bf 650}, 128 (2006), [\eprintmod[astro-ph/0603205]],
\newblock {(KVN)}.

\bibitem{2009ApJ...700..989B}
P.~{Bode}, J.~P. {Ostriker} and A.~{Vikhlinin},
\newblock \apj {\bf 700}, 989 (2009), [\eprintmod[0905.3748]].

\bibitem{TinkerEtal}
J.~L. {Tinker} {\em et~al.},
\newblock ArXiv e-prints {\bf 803} (2008), [\eprintmod[0803.2706]].

\bibitem{Kometal08}
E.~{Komatsu} {\em et~al.},
\newblock ArXiv e-prints {\bf 803} (2008), [\eprintmod[0803.0547]].

\bibitem{Hu:2003pt}
W.~Hu and B.~Jain,
\newblock Phys. Rev. {\bf D70}, 043009 (2004), [\eprintmod[astro-ph/0312395]].

\bibitem{Riess:2006fw}
A.~G. Riess {\em et~al.},
\newblock Astrophys. J. {\bf 659}, 98 (2007), [\eprintmod[astro-ph/0611572]].

\bibitem{Astier:2005qq}
The SNLS, P.~Astier {\em et~al.},
\newblock Astron. Astrophys. {\bf 447}, 31 (2006),
  [\eprintmod[astro-ph/0510447]].

\bibitem{WoodVasey:2007jb}
ESSENCE, W.~M. Wood-Vasey {\em et~al.},
\newblock Astrophys. J. {\bf 666}, 694 (2007), [\eprintmod[astro-ph/0701041]].

\bibitem{Davis:2007na}
T.~M. Davis {\em et~al.},
\newblock Astrophys. J. {\bf 666}, 716 (2007), [\eprintmod[astro-ph/0701510]].

\bibitem{RiessEtal09}
A.~G. {Riess} {\em et~al.},
\newblock \apj {\bf 699}, 539 (2009), [\eprintmod[0905.0695]].

\bibitem{Komatsu09}
E.~{Komatsu} {\em et~al.},
\newblock \apjs {\bf 180}, 330 (2009), [\eprintmod[0803.0547]].

\bibitem{PercivalEtal09}
W.~J. {Percival} {\em et~al.},
\newblock ArXiv e-prints  (2009), [\eprintmod[0907.1660]].

\bibitem{ApplebyBattye}
S.~{Appleby} and R.~{Battye},
\newblock Physics Letters B {\bf 654}, 7 (2007), [\eprintmod[0705.3199]].

\bibitem{Starobinsky07}
A.~A. {Starobinsky},
\newblock Soviet Journal of Experimental and Theoretical Physics Letters {\bf
  86}, 157 (2007), [\eprintmod[0706.2041]].

\bibitem{NojOdin07}
S.~{Nojiri} and S.~D. {Odintsov},
\newblock Physics Letters B {\bf 657}, 238 (2007), [\eprintmod[0707.1941]].

\bibitem{KW}
J.~{Khoury} and M.~{Wyman},
\newblock ArXiv e-prints  (2009), [\eprintmod[0903.1292]].

\bibitem{Schmidt09}
F.~Schmidt,
\newblock \prd {\bf 80}, 043001 (2009), [\eprintmod[0905.0858]].

\end{thebibliography}

\end{document}